\documentclass[11pt]{article}

\usepackage{amsmath,amssymb,amsfonts,latexsym,amscd,amsthm}
\usepackage[colorlinks=true,plainpages=false,linktocpage,linkcolor=blue,citecolor=green,pagebackref]{hyperref}
\usepackage[letterpaper,left=1.5in,top=1in,bottom=1in,right=1.5in,includeheadfoot,headheight=16.6pt]{geometry}
\usepackage{graphicx}
\begin{document}

\title{Mini-maximizing two qubit quantum computations}
\author{\Large{Faisal Shah Khan\footnote{Department of Applied Mathematics and Sciences, Email: faisal.khan@kustar.ac.ae}, Simon J.D. Phoenix\footnote{Department of Electrical \& Computer Engineering, Email: simon.phoenix@kustar.ac.ae}} \\ \\
Khalifa University, Abu Dhabi, UAE}
\maketitle

\begin{abstract}
Two qubit quantum computations are viewed as two player, strictly competitive games and a game-theoretic measure of optimality of these computations is developed. To this end, the geometry of Hilbert space of quantum computations is used to establish the equivalence of game-theoretic solution concepts of Nash equilibrium and mini-max outcomes in games of this type, and quantum mechanisms are designed for realizing these mini-max outcomes. 
\end{abstract}

\section{Introduction}

A central feature of Shannon's \cite{Shannon} theory of information is the notion of probability distribution, that is, a convex linear combination of outcomes of some experiment. Replacing probability distributions with quantum superpositions, that is, complex projective linear combinations of outcomes of the experiment, and then following by measurement has important consequences for the treatment of information in the quantum realm. The resultant generalization of Shannon theory of information underpins many of the important applications of quantum mechanics, such as quantum computing, and the much broader theory of quantum information processing that embeds the classical treatments of probability distributions applied to observables within that of positive-definite, trace preserving maps.

Much of the work on quantum games to date has focused on the issue of what quantum mechanics brings to game theory. The perspective is decidedly that quantum mechanics can offer some potential advantage, such as being able to resolve game-theoretic conundrums (one such being that of the Prisoner's Dilemma), by the utilization of quantum resources unavailable within a
classical framework of the game. Whilst this focus is interesting, it is our belief that game-theory is at its most fruitful when applied as an analysis tool for situations in which there may be an underlying competitive element. Game theory has found such fruitful applications in fields as diverse as Economics and Biology. Here we are interested in whether the tools of game theory will allow us to understand, or design, quantum physical processes in a different way. In our previous work in \cite{Khan} we have characterized this approach as that of ``gaming the quantum'' rather than quantizing a game \cite{Bleiler}.

Quantum computation has been shown to behave differently than classical computation. Well known examples of this different behavior of quantum computation include Grover's quantum search algorithm \cite{Grover} for identifying an element from a list which outperforms its classical counterpart by a factor that is quadratic in the size of the list, and Shor's \cite{Shor} factoring algorithm that offers an exponential speed up over classical algorithms for factoring large integers. Meyer \cite{Meyer} has posed the question whether game theory can be used to gain a different insight into quantum computation and, in particular, whether the notion of Nash equilibrium applied to quantum computation might be a useful tool in the design of quantum algorithms. With this in mind, a game-theoretic treatment of two qubit quantum computations as two player, strictly competitive games is presented here in section
\ref{sec:quantumgames}. In particular, we identify Nash equilibrium outcomes in these quantum games as characterized by the geometry of the Hilbert space of quantum computations, an idea we first presented in \cite{Khan}. Expanding on this earlier work, we show here that this geometric characterization implies the equivalence of Nash equilibrium and mini-max outcomes in quantum games of this type. Further, in section \ref{sec:mechanism}, we design quantum mechanisms that identifies mini-max outcomes in two qubit quantum
computations via linear programming. 

This approach allows us to consider the output of some quantum process as optimal in the sense of Nash equilibrium. In the case of two qubit quantum computations considered here, a computation of this type is said to be {\it optimal} if the output it produces is the best possible under the constraints given. The conditions for reverse-engineering such a quantum computation are laid out in generality by the analysis of the mini-max conditions we derive. We give the example of the implementation of a CNOT gate in section \ref{sec:CNOT} to illustrate this principle. This perspective also has ramifications when considering the evolution of quantum states in general. If we consider the maximally entangled state as the output of some quantum game in which the players generate the entangled state as an equilibrium output then we can understand maximal entanglement as a consequence of the ``push-pull'' between the players competing strategies. Our reverse-engineering approach allows us to construct the family of games in section \ref{sec:EntanglementGame} for which a maximally entangled state is the mini-max output state produced and we illustrate this with an example.

We end this section with an open question: can Grover's quantum search type algorithms be cast as strictly competitive quantum games so that the resulting game model offers insights into quantum informational processes? 

\section{Strictly competitive quantum games and Nash equilibrium}\label{sec:quantumgames}

A finite non-cooperative quantum game $\Phi$ is a function with a $d$-dimensional Hilbert space ${\rm {\bf H}}_d$ of quantum superpositions as its range, together with the additional feature that ``players'' entertain non-identical preferences over the elements of the range. In symbols
\begin{equation}\label{qgame}
\Phi: \prod_i D_i \longrightarrow {\rm {\bf H}}_d
\end{equation}
The factor $D_i$ in the domain of $\Phi$ is the set of {\it quantum strategies} of player $i$ and a {\it play} of the game $\Phi$ is a tuple of strategies in $\prod_i D_i$ producing a {\it payoff} to each player in the form of an {\it outcome}, that is, an element of the range of $\Phi$. If upon some appropriate restrictions $\Phi$ reduces to some non-quantum ``classical'' game $\Gamma$, then $\Phi$ is said to be a {\it quantization} of $\Gamma$. As Bleiler points out in \cite{Bleiler}, such quantized games can be viewed as extensions of some classical game $\Gamma$ into the quantum realm in the same way a {\it mixed} game is an extension of a game into the realm of probabilities. For a detailed treatment of game theory, readers are referred to Binmore's excellent book on the subject \cite{Binmore}.

From now on, the terms quantum strategy and strategy will be used interchangeably. A {\it Nash equilibrium} is a play of a (quantum) game in which every player employs a strategy that is a best reply to the strategic choice of every other players. In other words, unilateral deviation from a Nash equilibrium by a player in the form of a different choice of strategy will produce an outcome that is less than or equal to in preference to {\it that} player than before. In \cite{Khan}, we characterize the notion of Nash equilibrium via the inner-product of the Hilbert space of quantum superpositions in a two player, {\it strictly competitive} quantum game, that is, a game in which the preferences of the players are diametrically opposite. Because of this diametrically opposite nature of the players' preferences, Nash equilibria in two player, strictly competitive games are plays in which each player minimizes his opponent's maximum possible payoff thus maximizing his own minimum possible payoff. This feature of Nash equilibria  in strictly competitive games is known as the mini-max theorem.

More precisely, if $\Phi$ is a two player strictly competitive quantum game, then 
\begin{equation}\label{sqgame}
\Phi: D_1 \times D_2 \longrightarrow {\rm {\bf H}}_4
\end{equation}
and the players' preferences may be defined as follows. Let $B=\left\{b_1,b_2,b_3,b_4 \right\}$ be an orthogonal basis of ${\rm {\bf H}}_4$ corresponding to some observable. Define the preferences of Player I over the elements of $B$ to be 
\begin{equation}\label{eqn:pref I}
b_1 \succ b_2 \equiv b_3 \equiv b_4
\end{equation}
and Player II's preferences to be 
\begin{equation}\label{eqn:pref II}
b_2 \succ b_1 \equiv b_3 \equiv b_4
\end{equation}
where the symbol $\succ$ stands for ``strictly preferred over'' and the symbols $\equiv$ stands for ``indifferent between''. Note the diametrically opposite nature of the players' preferences with respect to the elements $b_1$ and $b_2$. These preferences of the players over the elements of the basis $B$ induce preferences over arbitrary quantum superpositions in ${\rm {\bf H}}_4$ via the notion of distance or angle between quantum superpositions. To be more precise, let $p$ and $q$ be quantum superpositions in ${\rm {\bf H}}_4$, and let $\theta_{(p,q)}$ denote the distance between the two quantum superpositions as measured by the angle between them. Then for Player I
\begin{equation}\label{qprefp}
q \succ p \quad {\rm whenever} \quad \theta_{(q,b_1)} < \theta_{(p,b_1)}.
\end{equation}
Similarly, if $r$ and $s$ are two quantum superpositions in ${\rm {\bf H}}_4$, then for Player II
\begin{equation}\label{rprefs}
r \succ s \quad {\rm whenever} \quad \theta_{(r,b_2)} < \theta_{(s,b_2)}.
\end{equation}
A play $(x^*, y^*)$ of the quantum game $\Phi$ is a Nash equilibrium if unilateral deviation by any one player from $(x^*, y^*)$ will produce a quantum superposition that is either equal to in distance or further away from that player's most preferred element of $B$. That is, if Player I unilaterally deviates from the $(x^*, y^*)$ and instead engages in any other play $(x, y^*)$ then
\begin{equation}\label{eqn:1}
\theta_{(\Phi(x,y^*),b_1)} \geq \theta_{(\Phi(x^*,y^*),b_1)}.
\end{equation}
Also, if Player II unilaterally deviates from $(x^*,y^*)$ and instead engages in any other play $(x^*,y)$ then
\begin{equation}\label{eqn:2}
\theta_{(\Phi(x^*,y),b_2)} \geq \theta_{(\Phi(x^*,y^*),b_2)}
\end{equation}
Inequalities (\ref{eqn:1}) and (\ref{eqn:2}) characterize a Nash equilibrium outcome in a two player, strictly competitive quantum game as a simultaneous minimization problem. Note however that this simultaneous minimization problem is equivalent to the one in which each player minimizes the maximum possible payoff, in terms of the quantity $\theta_{(,)}$, of the other player. This is further equivalent to the problem in which each player maximizes his own minimum payoff in terms of the quantity $\theta_{(,)}$. We therefore have the following classic result from two player, strictly competitive game theory for quantum games of the same type. \\

\noindent {\bf Proposition}: {\it An outcome of a two player, strictly competitive quantum game $\Phi$ is a Nash equilibrium if and only if it is a mini-max (or maxi-min) outcome.} \\

In the broader context of quantum information, the strictly competitive, quantum game $\Phi$ described above can be viewed as an information channel processing quantum information with outcomes of this channel marked with respect to non-identical preferences of some background players. Whether this game-theoretic perspective and the notion of Nash equilibrium leads to any new insights or results in the area of quantum information processing is an open question; indeed, it is an undiscovered country. 

Leaving the exploration of this undiscovered country to a subsequent expedition, we restrict here only to the much simpler case of quantum computation with two qubits. We point out that {\it qubit} is simply another term for quantum superpositions defined for two possible outcomes of some experiment. As dictated by the axioms of quantum mechanics, the strictly competitive quantum game is now necessarily a unitary operator 
\begin{equation}\label{unitarygame}
U: H_2 \times H_2 \longrightarrow H_2 \otimes H_2
\end{equation}
where $D_1=D_2=H_2$ is the Hilbert space of a qubit designated here as each player's set of quantum strategies, and $H_2 \otimes H_2$, the Hilbert space of the {\it joint states} of two qubits, is designated as the set of outcomes. At this point, the theory of Hilbert space guarantees that the image of the unitary operator $U$ is a sub-Hilbert space of $H_2 \otimes H_2$. Moreover, for a given sub-Hilbert space $\mathcal{S}$ of a Hilbert space $\mathcal{H}$, there exists a unique element $s \in \mathcal{S}$ that minimizes the ``distance'' between elements of $\mathcal{S}$  and a fixed $h \in \mathcal{H}$ that can be measured by $\theta_{(s,h)}$. 

Therefore, with $\mathcal{H}=H_2 \otimes H_2$, and ${\rm Image}(U)=\mathcal{S}$, it is guaranteed that there exist $s_{m_1}, s_{m_2} \in \mathcal{S}$ that minimize the distance between $b_1, b_2\in \mathcal{H}$, respectively. We seek here a {\it mini-maximizer}, that is, $s_m \in \mathcal{H}$ such that $s_m=s_{m_1}=s_{m_2}$. In the following section, we take a mechanism design approach to construct a two player, strictly competitive quantum game $U$ and the strategic choices of the players in $U$ that realize mini-maximizers. 

\section{Mechanism design for two player, strictly competitive quantum games at mini-max}\label{sec:mechanism}

In this section, we take the quantum game $U$ to be a generic $4 \times 4$ unitary matrix or, in other words, a generic two qubit quantum logic gate, and explore features of this game and the quantum strategic choices that the players will make in this game in order to observe a mini-max outcome. Let
\begin{equation}\label{basis}
B=\left\{ b_1=\left|0\right\rangle \otimes \left| 0 \right\rangle,b_2=\left|0 \right\rangle \otimes \left|1 \right\rangle,b_3=\left|1 \right\rangle \otimes \left|0 \right\rangle,b_4=\left|1 \right\rangle \otimes \left| 1 \right\rangle \right\},
\end{equation}
be the standard (orthogonal) computational basis for $H_2 \otimes H_2$, with 
\begin{equation}\label{kettovector}
\left|0 \right\rangle = \left(\begin{array}{c} 
1 \\
0\end{array}\right); \quad \left| 1 \right\rangle = \left(\begin{array}{c} 
0 \\
1\end{array}\right) 
\end{equation}
and preferences of the players defined as in expressions (\ref{eqn:pref I}) and (\ref{eqn:pref II}) and let 
\begin{equation}\label{unitarymatrix}
U=\left(\begin{array}{cccc}
U_{11} & U_{12} & U_{13} & U_{14} \\ 
U_{21} & U_{22} & U_{23} & U_{24}\\
U_{31} & U_{32} & U_{33} & U_{34} \\
U_{41} & U_{42} & U_{43} & U_{44}
\end{array}\right)
\end{equation}
be a strictly competitive, two player, two strategy quantum game represented with respect to the basis $B$. Also, let $B'=\left\{\left|0 \right\rangle, \left|1 \right\rangle \right\}$ be the computational basis for each $H_2$ that constitutes the players' set of strategies. A strategy for Player I is the choice of a qubit state, say
\begin{equation}\label{PlayerIstrategy}
A=x_1\left|0 \right\rangle+y_1\left|1 \right\rangle
\end{equation}
and a strategy for Player II is the choice of qubit state, say 
\begin{equation}\label{PlayerIIstrategy}
B=x_2\left|0 \right\rangle+y_2\left|1 \right\rangle.
\end{equation}
Note that these quantum strategic choices can be physically implemented via local unitary operations
\begin{equation}\label{eqn:PIStrat}
\left(\begin{array}{cc}
x_1 & -\overline{y}_1\\ 
y_1 & \overline{x}_1\\
\end{array}\right); \quad |x_1|^2+|y_1|^2=1
\end{equation}
and
\begin{equation}\label{eqn:PIIStrat}
\left(\begin{array}{cc}
x_2 & -\overline{y}_2\\ 
y_2 & \overline{x}_2\\
\end{array}\right); \quad |x_2|^2+|y_2|^2=1
\end{equation}
on the qubit state $\left|0 \right\rangle$ respectively. Also note that the notion of quantum strategy depends on the ``choice'' of some initial state on which players act with appropriate unitary operators. This feature of quantum strategies, as defined here, of acting on some initial state arises from the underlying physics of the quantum system being gamed and is an idea not typically entertained in the toy game models used in game theory.  

Next, consider a mini-max play of the quantum game $U$, that is, a play in which Player I chooses the strategy
\begin{equation}\label{minimaxstratI}
A^*=\left(\begin{array}{c} 
x_1^* \\
y_1^*\end{array}\right)
\end{equation}
and Player II chooses the strategy
\begin{equation}\label{minimaxstratII}
B^*=\left(\begin{array}{c} 
x_2^* \\
y_2^* \end{array}\right).
\end{equation}
The output of the game is then
\begin{equation}\label{eqn:output}
U(A^*,B^*)=s_m=\left(\begin{array}{c}
U_{11}x^*_1x^*_2+U_{12}x^*_1y^*_2+U_{13}y^*_1x^*_2+U_{14}y^*_1y^*_2\\ 
U_{21}x^*_1x^*_2+U_{22}x^*_1y^*_2+U_{23}y^*_1x^*_2+U_{24}y^*_1y^*_2\\
U_{31}x^*_1x^*_2+U_{32}x^*_1y^*_2+U_{33}y^*_1x^*_2+U_{34}y^*_1y^*_2\\ 
U_{41}x^*_1x^*_2+U_{42}x^*_1y^*_2+U_{43}y^*_1x^*_2+U_{44}y^*_1y^*_2
\end{array}\right)
\end{equation}
with $\left\| s_m \right\|^2=1$ and 
\begin{equation}\label{angle1}
\theta_{(s_m,b_1)}=\cos^{-1}(\left| U_{11}x^*_1x^*_2+U_{12}x^*_1y^*_2+U_{13}y^*_1x^*_2+U_{14}y^*_1y^*_2 \right|^2)
\end{equation}
\begin{equation}\label{angle2}
\theta_{(s_m,b_2)}=\cos^{-1}(\left| U_{21}x^*_1x^*_2+U_{22}x^*_1y^*_2+U_{23}y^*_1x^*_2+U_{24}y^*_1y^*_2 \right|^2).
\end{equation}

\subsection{Mini-maximizer}\label{sec:minimax}

A unilateral deviation on part of Player I from the play $(A^*, B^*)$ to {\it any} other $(A,B^*)$ will produce the output state $\widehat{s}$ such that 
\begin{equation}\label{deviationI}
\theta_{(\widehat{s},b_1)}=\cos^{-1}(\left| U_{11}x_1x^*_2+U_{12}x_1y^*_2+U_{13}y_1x^*_2+U_{14}y_1y^*_2 \right|^2)
\end{equation}
with 
\begin{equation}\label{compareangleI}
\theta_{(\widehat{s},b_1)} \geq \theta_{(s_m,b_1)}.
\end{equation}
Because the inverse cosine function is decreasing, it follows that 
\begin{equation}\label{inequality1}
\left| U_{11}x_1x^*_2+U_{12}x_1y^*_2+U_{13}y_1x^*_2+U_{14}y_1y^*_2 \right|^2 \leq \left| U_{11}x^*_1x^*_2+U_{12}x^*_1y^*_2+U_{13}y^*_1x^*_2+U_{14}y^*_1y^*_2 \right|^2.
\end{equation}
Furthermore, because the quadratic function is one-to-one and increasing for non-negative inputs, it follows that  
\begin{equation}\label{eqn:ineq1}
\left| U_{11}x_1x^*_2+U_{12}x_1y^*_2+U_{13}y_1x^*_2+U_{14}y_1y^*_2 \right| \leq \left| U_{11}x^*_1x^*_2+U_{12}x^*_1y^*_2+U_{13}y^*_1x^*_2+U_{14}y^*_1y^*_2 \right|.
\end{equation}

Likewise, a unilateral deviation on part of Player II from the play $(A^*, B^*)$ to {\it any} other $(A^*,B)$ will produce the output state $\widetilde{s}$ such that
\begin{equation}\label{deviationII}
\theta_{(\widetilde{s},b_2)}=\cos^{-1}(\left| U_{21}x^*_1x_2+U_{22}x^*_1y_2+U_{23}y^*_1x_2+U_{24}y^*_1y_2 \right|^2)
\end{equation}
such that 
\begin{equation}\label{eqn:ineq2}
\left| U_{21}x^*_1x_2+U_{22}x^*_1y_2+U_{23}y^*_1x_2+U_{24}y^*_1y_2 \right| \leq \left| U_{21}x^*_1x^*_2+U_{22}x^*_1y^*_2+U_{23}y^*_1x^*_2+U_{24}y^*_1y^*_2 \right|.
\end{equation}
Factoring and applying triangle inequality on the right-hand side of inequality (\ref{eqn:ineq1}) reduces it to
\begin{equation}\label{triangleineq1}
\left| U_{11}x_1x^*_2+U_{12}x_1y^*_2+U_{13}y_1x^*_2+U_{14}y_1y^*_2 \right| \leq \left| U_{11}x_2^*+U_{12}y_2^*\right|\cdot \left| x_{1}^*\right|+\left| U_{13}x_2^*+U_{14}y_2^*\right|\cdot\left| y_{1}^*\right|
\end{equation}
which can be expressed compactly as
\begin{equation}\label{eqn:shortineq}
\left| U_{11}x_1x^*_2+U_{12}x_1y^*_2+U_{13}y_1x^*_2+U_{14}y_1y^*_2 \right| \leq P\left| x_{1}^* \right| + Q \left| y_{1}^* \right| 
\end{equation}
with $P=\left| U_{11}x_2^*+U_{12}y_2^*\right|$ and $Q=\left| U_{13}x_2^*+U_{14}y_2^*\right|$. Applying a similar reasoning to inequality (\ref{eqn:ineq2}) produces
\begin{equation}\label{eqn:shortineq1}
\left| U_{21}x^*_1x_2+U_{22}x^*_1y_2+U_{23}y^*_1x_2+U_{24}y^*_1y_2 \right| \leq P'\left| x_{2}^* \right| + Q' \left| y_{2}^* \right|. 
\end{equation}
with $P'=\left| U_{21}x_1^*+U_{23}y_1^*\right|$ and $Q'=\left| U_{22}x_1^*+U_{24}y_1^*\right|$. Inequalities (\ref{eqn:shortineq}) and (\ref{eqn:shortineq1}) can be further simplified by factoring and applying the triangle inequality to their respective left-hand sides, giving two cases each:
\begin{equation}\label{eqn:case1}
P|x_1| +Q|y_1| \leq P| x_{1}^*| + Q| y_{1}^*|
\end{equation} 
or
\begin{equation}\label{eqn:case2}
P|x_1| +Q|y_1| \geq P| x_{1}^*| + Q| y_{1}^*|
\end{equation} 
and 
\begin{equation}\label{eqn:case3}
P'|x_2| +Q'|y_2| \leq P'| x_{2}^*| + Q'| y_{2}^*|
\end{equation} 
or 
\begin{equation}\label{eqn:case4}
P'|x_2| +Q'|y_2| \geq P'| x_{2}^*| + Q'| y_{2}^*|.
\end{equation} 
The quantum game $U$ will entertain a mini-max outcome when inequalities (\ref{eqn:case1}) and (\ref{eqn:case3}) are satisfied, or (\ref{eqn:case1}) and (\ref{eqn:case4}), or (\ref{eqn:case2}) and (\ref{eqn:case3}), or (\ref{eqn:case2}) and  (\ref{eqn:case4}) are satisfied.  

To analyze the first case of mini-maximizer as determined by inequalities (\ref{eqn:case1}) and (\ref{eqn:case3}), first rewrite inequality (\ref{eqn:case1}) as
\begin{equation}\label{eqn:rewrite}
\left( \left| x_1 \right| + \frac{Q}{P} \left| y_1 \right| \right) -  \frac{Q}{P}\left| y_1^* \right| \leq \left| x_1^* \right|
\end{equation}
when $P \neq 0$. Geometrically, the family of solutions of inequality (\ref{eqn:rewrite}) consists of all those sets of points in $\mathbb{R}^2$ such that the points are on or above the line $\left( \left| x_1 \right| + \frac{Q}{P} \left| y_1 \right| \right) -  \frac{Q}{P}\left| y_1^* \right| = \left| x_1^* \right|$, with $|y_1^*|$ as the independent variable and $|x_1^*|$ as the dependent variable, and which is also confined to lie within the unit circle $|x_1^*|^2+|y_1^*|^2=1$. The condition $P \neq 0$ necessary to derive inequality (\ref{eqn:rewrite}) can be made game-theoretically meaningful and quantum mechanically insightful by viewing it as an artifact of the quantum game $U$ that manifests via appropriate conditions on the quantities $U_{11}$ and $U_{12}$, in contrast to the view that it is an artifact of special assumptions regarding the parameters $|x_2^*|$ or $|y_2^*|$ that define Player II's choice of strategy at mini-max. This view keeps the mini-max analysis as general as possible.

A similar analysis applies to Player II's strategic choice at mini-max via the inequality 
\begin{equation}\label{eqn:rewrite1}
\left( \left| x_2 \right| + \frac{Q'}{P'}\left| y_2 \right| \right) - \frac{Q'}{P'} \left| y_2^* \right| \leq \left| x_2^* \right|
\end{equation}
derived from inequality (\ref{eqn:case3}) when $P'\neq 0$. As before, geometrically, the family of solutions of inequality (\ref{eqn:rewrite1}) consists of all those sets of points in $\mathbb{R}^2$ such that the points are on or above the line $\left( \left| x_2 \right| + \frac{Q'}{P'}\left| y_2 \right| \right) - \frac{Q'}{P'} \left| y_2^* \right| =\left| x_2^* \right|$, with $|y_2^*|$ as the independent variable and $|x_2^*|$ as the dependent variable, and which is also confined to lie within the unit circle $|x_2^*|^2+|y_2^*|^2=1$. As was done above, the condition $P' \neq 0$ necessary to derive inequality (\ref{eqn:rewrite1}) can be made game-theoretically meaningful and quantum mechanically insightful by viewing it as an artifact of the quantum game $U$ that manifests via appropriate conditions on the quantities $U_{11}$ and $U_{12}$, in contrast to the view that it is an artifact of special assumptions regarding the parameters $|x_1^*|$ or $|y_1^*|$ that define Player I's choice of strategy at mini-max. Again, this view keeps the mini-max analysis as general as possible.

\begin{figure}
\centering
\includegraphics[scale=0.5]{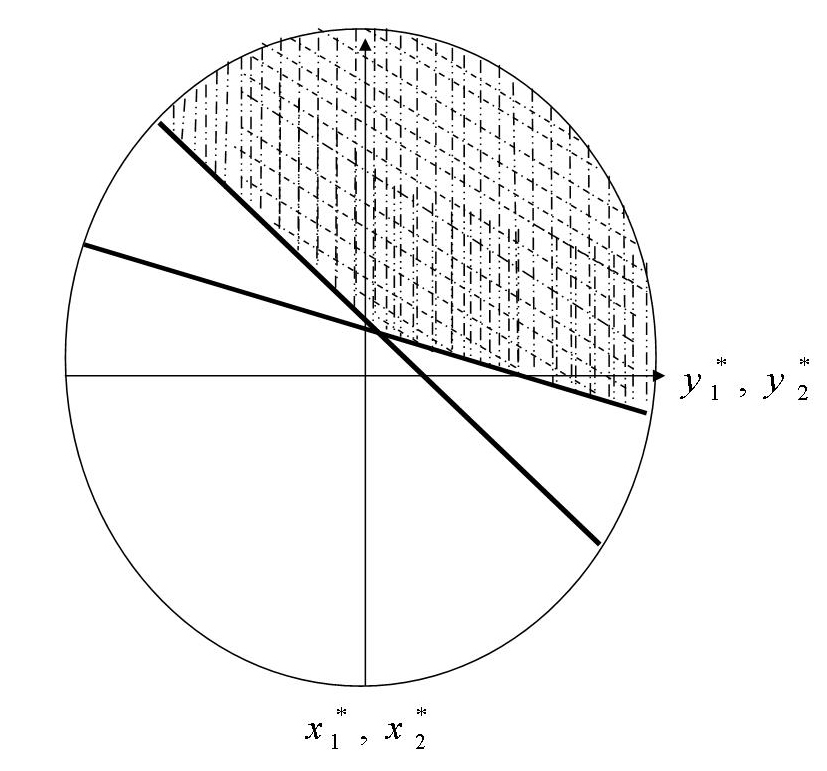}
\caption{\small{Region in $\mathbb{R}^2$ corresponding to mini-max outcomes in two player, strictly competitive quantum games satisfying $P \neq 0$ and $P' \neq 0$.} }
\label{fig:NE1}
\end{figure}

Figure \ref{fig:NE1} captures in a generic sense the region that defines Nash equilibrium plays in two player, strictly competitive quantum games $U$ that satisfy the conditions $P \neq 0$ and $P' \neq 0$. The horizontal axis represents both $|y_1^*|$ and $|y_2^*|$ and the vertical axis represents both $|x_1^*|$ and $|x_2^*|$. A play $(A^*, B^*)$ of a strictly competitive two player, two strategy quantum game is a Nash equilibrium if the parameters $x_1^*$ and $y_1^*$ defining Player I's strategy $A^*$ and the parameters $x_2^*$ and $y_2^*$ defining Player II's strategy $B^*$ satisfy the constraints of Figure 1. 

Note that one could instead isolate the quantities $|y_1^*|$ and $|y_2^*|$ by assuming that $Q \neq 0$ and $Q' \neq 0$ in the quantum game $U$, or indeed isolate $|x_1^*|$ and $|y_2^*|$ when $P \neq 0$ and $Q' \neq 0$ and isolate $|y_1^*|$ and $|x_2^*|$ when $P' \neq 0$ and $Q \neq 0$. Any one of these cases will produce a region in $\mathbb{R}^2$ bounded by lines that will correspond to mini-max outcomes in those two player, strictly competitive quantum games $U$ that satisfy the conditions of the case.   
 
The remaining inequality pairs corresponding to mini-max outcomes can be analyzed similarly. 

\section{The CNOT quantum game}\label{sec:CNOT}

Consider the controlled-not (CNOT) quantum logic gate 
\begin{equation}\label{CNOT}
U=\left(\begin{array}{cccc}
1 & 0 & 0 & 0 \\ 
0 & 1 & 0 & 0 \\
0 & 0 & 0 & 1\\ 
0 & 0 & 1 & 0
\end{array}\right)
\end{equation}
as the two player, strictly competitive quantum game $U$ discussed above. In this case, $P=\left| x_2^*\right|$, $Q=0=P'$, $Q'=\left|x_1^*\right|$. Inequality (\ref{eqn:rewrite}) gives
\begin{equation}
\left| x_1\right| \leq \left| x_1^* \right|; 
\end{equation}
however, we cannot utilize inequality (\ref{eqn:rewrite1}) because $P'=0$. Instead, we choose to rewrite inequality (\ref{eqn:case3}) as 
\begin{equation}\label{eqn:rewrite3}
\left( \left| y_2 \right| + \frac{P'}{Q'}\left| x_2 \right| \right) - \frac{P'}{Q'} \left| x_2^* \right| \leq \left| y_2^* \right|
\end{equation}
assuming that $R'=\left| x_1^*\right| \neq 0$. This gives
\begin{equation}\label{CNOTcondition}
\left| y_2 \right| \leq \left| y_2^* \right|
\end{equation}
meaning that a mini-maximizer sees Player I employ a quantum strategy for which the length of the $x$-coordinate is the maximum possible and Player II employ a quantum strategy for which the length of the $y$-coordinate is maximum possible. This can be achieved by Player I choosing the component $\left|x^*_1\right|$ of his quantum strategy to be any complex number of length 1 and Player II choosing the component $\left| y^*_2\right|$ of his quantum strategy to be any complex number of length 1. In other words, for a mini-maximizer in the CNOT game based on the conditions of inequalities (\ref{eqn:rewrite}) and (\ref{eqn:rewrite3}) 
\begin{equation}\label{CNOTresult}
A^*=\alpha\left|0\right\rangle, \quad B^*=\beta\left|1\right\rangle; \quad \left| \alpha\right|^2=\left| \beta\right|^2=1, 
\end{equation}
that is, Player I chooses a quantum strategy from the subspace of $H_2 \otimes H_2$ generated by the state $\left|0\right\rangle$ whereas Player II chooses a quantum strategy from the subspace spanned by $\left|1\right\rangle$. 

Reverting back to quantum logic gate perspective, the preceding discussion tells us that the CNOT gate performs an optimal quantum computation, in the sense of mini-max, when its input states satisfy the conditions in (\ref{CNOTresult}). That is, the CNOT gate performs optimally exactly when the input states are separable, a result that is perhaps surprising given that entanglement between states is often expected to be the source of interesting quantum mechanical behavior. 

Mini-maximizers in the CNOT game based on the conditions specified by inequalities (\ref{eqn:case1}) and (\ref{eqn:case4}), (\ref{eqn:case2}) and (\ref{eqn:case3}), and (\ref{eqn:case2}) and (\ref{eqn:case4}) can be analyzed similarly. 

\section{Mechanism design for maximal entanglement}\label{sec:EntanglementGame}

Consider the maximally entangled state
\begin{equation}\label{maxent}
M=\frac{1}{\sqrt{2}}\left(\left|00\right\rangle + \left| 11 \right\rangle\right)
\end{equation}
as a mini-maximizer $s_m$ in a two player, strictly competitive game in which the corresponding strategic choice of each player is the qubit state $\left|0 \right\rangle$, and  preferences of the players are identified by (\ref{eqn:pref I}) and (\ref{eqn:pref II}). We design quantum mechanisms that will entertain $M$ as a mini-maximizer. Note that for $M$, the output state (\ref{eqn:output}) of the quantum mechanism $U$ requires that $U_{21}=0$, from which it follows that $P'=0$. In this situation, we cannot utilize inequality (\ref{eqn:rewrite1}) which requires the contrary. However, note that for $M$, $Q'$ need not be equal to $0$ and therefore by assuming that $Q' \neq 0$, we can use inequality (\ref{eqn:rewrite3}) here to get
\begin{equation}
\left|y_2\right| \leq 0 \Rightarrow y_2=0.
\end{equation}
This last result gives no insight into properties of the quantum mechanism $U$. However, inequality (\ref{eqn:rewrite}) gives
\begin{equation}\label{eqn:rewrite4}
\left| x_1\right| + \sqrt{2}\left|U_{13}\right| \left| y_1\right| \leq 1 \Leftrightarrow \left|U_{13} \right| \leq \frac{1}{\sqrt{2}\left|y_1 \right|}\left(1-\left|x_1\right| \right).
\end{equation}
Therefore, the parameter $U_{13}$ of the quantum game $U$ must satisfy inequality (\ref{eqn:rewrite4}) in terms of a strategic choice of player I {\it other} than the mini-maximizer. Switching to the quantum computational perspective, the preceding discussion tells us that a quantum mechanism $U$ will produce the maximally entangled state (\ref{maxent}) as a mini-maximizer, with ground state $\left|00 \right\rangle$ as the input, exactly when its component $U_{13}$ satisfies inequality (\ref{eqn:rewrite4}) for {\it any} value of one of the input qubits other than the ground state. 

Other quantum mechanisms can be constructed that take the strategic choices of the players corresponding to the ground state to the maximally entangled state considered here as a mini-maximizer, based on the other possible combinations of the inequalities introduced in section \ref{sec:minimax} as well as for other possible strictly competitive preferences of the players. 
 
\section{Discussion}

We set up a game-theoretic model for viewing two qubit quantum computations as strictly competitive quantum games and utilized a geometric characterization
of the notion of Nash -equilibrium in such quantum games developed in \cite{Khan} to classify Nash equilibria in the form of mini-max outcomes in one instance of these games via regions in $\mathbb{R}^{2}$, and set up conditions for the classification of mini-max outcomes in other instances. We propose that a mini-max outcome can be used as a measure of optimality of a quantum computation under strictly competitive constraints such as the one's discussed in section \ref{sec:quantumgames}.

Gaming the quantum is a natural course of action when exploring the quantum landscape for constrained optimization. We gamed only the relatively
straightforward case of two qubit quantum computations here, modeled as strictly competitive games with a specific choice of preferences for the
players; an extension to this is to investigate these games with different choices of preferences for the players by relaxing the strictly competitive
condition. Our initial results in this direction suggest that this is richer and more subtle than the strictly competitive case considered here. Further,
gaming of $n$ qubit and continuous variable quantum computations, initially as strictly competitive games and later as more general non-cooperative games,
and exploring the nature of Nash equilibria more general than mini-max in these computations is another important extension to the work we have
presented here. Even further, more general functions permitted in the quantum realm such as those that describe mixed quantum states, might be amenable to
gaming. In this context, the question of capacity of a quantum informational channel, a notion that has been described in the past as a mini-max problem in
a non-game theoretic context \cite{Schumacher, Cortese}, might be re-cast as a strictly competitive game. The emerging area of a quantum neural networks has
already been cast as a gaming problem with the notion of mini-max outcomes corresponding to that of optimal learning strategy for the network \cite{Khan1}.

Another important open question of a fundamental nature is whether the evolution of Schr\"odinger's equation might be gamed to obtain fresh insight into how one might speak meaningfully of ``optimality"' of various features of quantum mechanics.

\end{document}